\documentstyle[pre,aps,twocolumn,epsbox]{revtex}

\begin{document}

\draft
\preprint{Phys. Rev. E}
\title{Relationships between a roller and a dynamic pressure 
distribution in circular hydraulic jumps}

\author{Kensuke Yokoi$^{1,2}$ and Feng Xiao$^{3}$}
\address{
$^1$Division of Mathematics and Research
Institute for Electronic 
Science, Hokkaido University, Sapporo 060-0812, Japan \\
$^2$Computational Science Division, RIKEN
(The Institute of Physical and Chemical Research),
Wako 351-0198, Japan \\
$^3$Department of Energy Sciences, Tokyo Institute of Technology,
Yokohama 226-8502, Japan }
\date{To appear in Phys. Rev. E, Vol.61, Feb. 2000}
\maketitle
\begin{abstract}
We investigated numerically the relation between a roller and the 
pressure distribution to clarify the dynamics 
of the roller in circular hydraulic jumps.
We found that a roller which characterizes a type II jump is 
associated with two high pressure regions after the jump, while a 
type I jump (without the roller) is associated with only one high
pressure region.  
Our numerical results show that building up an appropriate pressure
field is essential for a roller.   

\end{abstract}

\begin{figure}[hbtp]
\begin{center}
\psbox[height=2.5in]{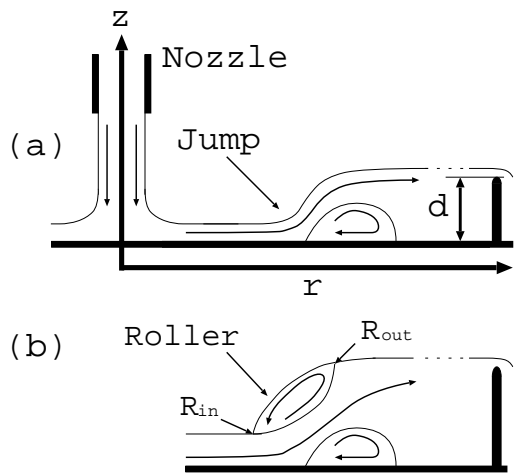}
\end{center}
\caption{Schematic figures of the circular hydraulic jump. The radius
         of the wall is much larger than the radius of the jump. The
         flow from the nozzle is constant. 
         In this experiment, a high viscous liquid is used for
         controlling the instability of flow pattern. 
         (a) and (b) are called type I and type II, respectively.
         The points of the inside and outside of the roller are
         defined as $(R_{in}^r,R_{in}^z)$ and
         $(R_{out}^r,R_{out}^z)$.}  
\label{fig:intoro1}
\end{figure}

As can be easily observed in a kitchen sink, a circular hydraulic jump 
is formed when a vertical liquid jet impinges on a horizontal surface.
The schematic figure of the circular hydraulic jump can be shown
as in Fig.\ \ref{fig:intoro1}.
The phenomenon has been investigated by many   
researchers through various approaches
\cite{rayleigh,tani,chow,hj8,hj7,hj6,hj2,hj5,hj4,hj9,ken1}.

In some experiments, the depth on the outside of
the jump can be controlled by varying the height of a circular wall
$d$, as shown in Fig.\ \ref{fig:intoro1}. 
Experimental results show that a circular hydraulic jump has two kinds 
of steady states which can be reached by changing $d$ \cite{hj2}. 
When $d$ is small or $0$, a type I jump is formed, as shown in
Fig.\ \ref{fig:intoro1}(a). 
On increasing $d$ the jump becomes steeper until a critical $d_c$ is 
reached. 
If $d$ becomes larger than $d_c$, the liquid outside of the jump
topples. Then another steady state, a type II jump, is formed as shown
in Fig.\ \ref{fig:intoro1}(b).
The eddy on the surface in a type II jump, a secondary circulation,
is usually called a ``roller.'' The existence
of a roller distinguishes the two types of jumps.   

The roller is a common and important feature for many
hydraulic  phenomena. 
Recent experiments\cite{hj4,hj9} demonstrate that various regular
polygonal jumps can develop  from a circular jump by controlling the
height of the outer circular wall, and that all those polygonal jumps
are associated with rollers.
Rollers are also observed in the channel flows and are useful for
dissipating the excess energy of high velocity flows, such as 
from sluice gates and spillways \cite{chow}.
It is widely recognized that rollers play an important role in
hydraulic engineering.

However, theoretical studies concerning the formation and
evolution of the  roller in a hydraulic jump are limited because of 
the largely deformed interface.
Some theoretical studies have been proposed using a
hydrostatic assumption in the vertical direction.
Some reasonable results have been obtained for the flows of type I
jumps \cite{hj5}. 
However, a type II jump appears to be beyond the regime
that this vertical-assumption
theoretical model is able to deal with.
Numerical modeling has also been used
to investigate the circular hydraulic jump problem.
Due to the difficulties in the numerical treatment of largely
distorted interfacial flows, the free boundary of the liquid surface
was treated as the fixed boundary of a prescribed shape \cite{hj2}.

\begin{figure}[hbtp]
  \begin{center}
  \psbox[height=1.33in]{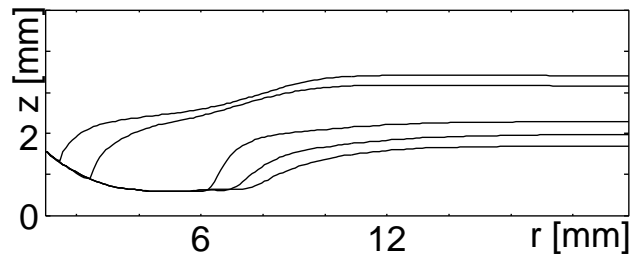}
  \end{center}
  \caption{Surface profiles for varying wall heights.
           $Q=5.6$ ml/s and $\nu_{l} = 7.6 \times 10^{-6}$ m$^2$/s are
           used. 
           The maximum values of the dynamic pressure
           around the jump (the primary high pressure) in type I jumps 
           were $1.77$, $3.99$, and $8.47$ Pa from the lowest
           respectively.} 
  \label{fig:result17}
\end{figure}

\begin{figure}[hbtp]
  \begin{center}
  \psbox[height=3.0in]{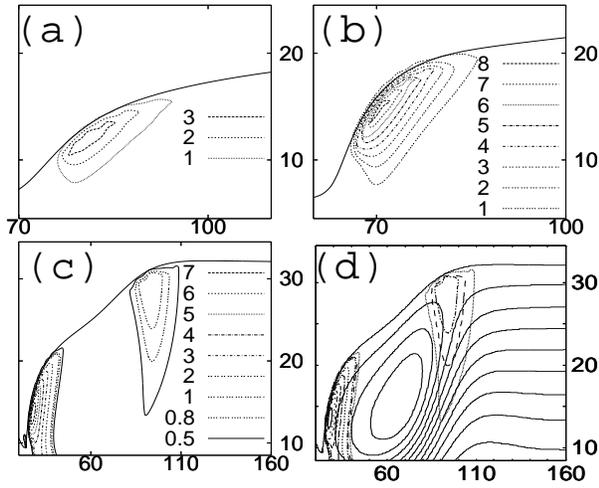}
  \end{center}
  \caption{Dynamic pressure (Pa) contours and the surface profiles 
           around the jump of the second (a),
           the third (b), and 
           the fourth (c) from the lowest in Fig.\ \ref{fig:result17}.
           (d) shows the streamline, the dynamic pressure contours, and
           the surface profile of the fourth profile.}   
  \label{fig:pp} 
\end{figure}

In our previous work \cite{ken1}, numerical simulations on circular
hydraulic jumps  were conducted using some newly developed numerical 
schemes for multi-fluid flows.
We investigated the transition from a type I jump to a type II jump. 
Non-hydrostatic pressure distributions in the gravitational direction
were observed in our simulations.  
In our studies, we call `dynamic pressure' the net amount of the
pressure resulting from extracting the hydrostatic pressure from the
actual pressure.  
We found that the dynamic pressure around the jump, which has
been neglected in most of the theoretical studies to date, is
important for the transition.  
In a type I jump, a steeper jump is always associated
with a higher wall height (\cite{hj2} and Fig.\ \ref{fig:result17}). 
Thus, as $d$ is increased, the curvature of the interface
immediately after the jump becomes larger, then
the surface tension is strengthened, 
because the surface tension is proportional to the curvature.
In order to counteract this surface tension and keep the jump surface 
steady, a larger rise in pressure is required
(Figs.\ref{fig:pp} (a,b)). 
If the wall height is increased over the critical $d_c$, the reverse
pressure gradient generated by the dynamic pressure becomes stronger
than the flow from below and a transition occurs. 

In this Rapid Communication, we intend to clarify the relationship
between the roller and the pressure field.
The simulation results show that the single high dynamic pressure
region in a type I jump becomes two regions after the transition to a
type II jump. 
These two high pressure regions are located along the
jump slope around the outer edge $R_{out}$ and the inner 
edge $R_{in}$ of the roller. 
This pressure distribution appears important to the flow
separation at the outer edge of the roller and then essential to the
maintenance of a roller.

The governing equations, including effects of gravity, viscosity and
surface tension can be written as
\begin{equation}
\frac{\partial \rho}{\partial t}
  + ({\bf u} \cdot \nabla) \rho = - \rho \nabla \cdot {\bf u},
\label{eq:onp1.1}
\end{equation}
\begin{equation}
\frac{\partial {\bf u}}{\partial t}
  + ({\bf u} \cdot \nabla) {\bf u} =
  - \frac{\nabla p}{\rho} + {\bf g}
  + \frac{\mu}{\rho} \Delta {\bf u}
  + \frac{\bf F_{sv}}{\rho},
\label{eq:onp1.2}
\end{equation}
\begin{equation}
\frac{\partial e}{\partial t}
  + ({\bf u} \cdot \nabla) e = - \frac{p}{\rho} \nabla \cdot {\bf u},
\label{eq:onp1.3}
\end{equation}
where $\rho$ is the density, ${\bf u}$ the velocity, p the pressure,
${\bf g}$ the gravitational acceleration,
$\mu$ the viscosity coefficient,
${\bf F_{sv}}$ the surface tension force,
and $e$ the inner energy.
Both the liquid and the gas are assumed to have an equation of state 
in the form of a polytropic gas, but with quite different sound
speeds (large for the liquid phase).

The numerical  model is constructed based on the 
C-CUP (CIP-Combined, Unified Procedure) method \cite{ccup}, 
the level set method \cite{level_set1,level_set2} and the CSF
(Continuum Surface Force) model \cite{csf}. 
By using the C-CUP method to  solve  multi-fluid flows,
we are able to deal with both the gas and 
the liquid phase in a unified framework,
and explicit treatment of the free boundary and interfacial 
discontinuity is not needed.

The interface between the liquid and the gas is tracked using the
level set method with the CIP (Cubic Interpolated Propagation) method
\cite{cip3} as the advection solver.
A density function $\phi$ generated from the level set function of the
level set method by the Heaviside function can be set as $\phi=1$ for
the liquid and $\phi=0$ for the air.  
The density function is then used to define the physical properties,
such as sound speed and viscosity for different materials. 

The surface tension force is modeled as a body force 
${\bf F_{sv}}$ calculated by the gradient of the density function,
${\bf F_{sv}} = \sigma \kappa \nabla \phi $,
where $\sigma$ is the fluid surface tension coefficient and $\kappa$ 
the local mean curvature.
$\kappa$ is computed from
$\kappa = - (\nabla \cdot {\bf n}),$
where ${\bf n}$ is the outgoing unit normal vector to the interface
and is evaluated from the level set function \cite{level_set2}.

\begin{figure}[hbtp]
  \begin{center}
  \psbox[height=1.33in]{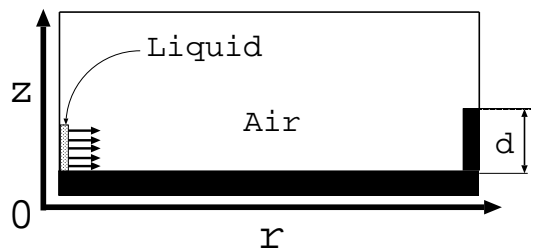}
  \end{center}
  \caption{Schematic figure for the initial condition of
          the simulation. 
          The dark part indicates the no-slip wall.
          The liquid is jetted from the lower left to the right
          direction.
          A Cartesian  
          grid with $\Delta r = \Delta z = 0.1$ mm is used.} 
  \label{fig:gairyaku}
\end{figure}

An axis-symmetric model has been constructed to deal with the circular 
hydraulic jump.
The configuration of the simulation model on an r-z plane is shown in  
Fig.\ \ref{fig:gairyaku}.
This calculation model is validated by comparing the computed results 
with the scaling relation \cite{hj6,ken1}.

Simulations were carried out with different heights of the outer
circular wall $d$.  
The volume flux of the inflow is $Q=5.6$ ml/s and the viscosity of the
liquid is $\nu_{l} = 7.6 \times 10^{-6}$ m$^2$/s.
The steady surface profiles for the various wall heights are
shown in Fig.\ \ref{fig:result17}.
The three lower profiles are type I jumps, and the two upper
profiles are type II jumps. 
We observe that the jump becomes steeper as the wall height increases
for a type I jump, while for a type II, the slope of the jump
appears less steep than that of  type I with a high wall hight. 
These are consistent with the experimental results \cite{hj2}.
The roller is usually a consequence of a steepened jump, while its 
occurrence always leads to the destruction of the steepness.  

The dynamic pressure distributions of the second, the third and the
fourth profiles from the lowest were plotted in
Figs.\ \ref{fig:pp}(a-c).
For the cases of type I (Figs.\ \ref{fig:pp}(a,b)), a high pressure
region 
(referred to hereafter as the primary high pressure) dominating a wide 
region under the jump surface is observed.
In a type II jump, two high pressure regions are developed around the
inner side of the jump (referred to as the primary high pressure)
and the outer side of the jump (refereed to as the secondary high
pressure) as shown in Fig.\ \ref{fig:pp}(c). 
This pressure distribution is essential for the roller.
We observe that the high pressure on the outer side
of the jump (the secondary high pressure) coincides with the
separation point of the flow, as shown in Fig.\ \ref{fig:pp}(d).
This secondary high pressure continuously provides a pressure gradient
force to maintain the upper reverse flow for the roller.
The secondary high pressure is associated with the surface tension.
In the steady state of the type II, the liquid surface appears
convexly curved around the secondary high pressure region or the outer
edge of the roller (Figs.\ \ref{fig:pp}(c,d)). 
This feature of the free surface around the outer edge of the roller
is also observed in experiment \cite{hj2}.
To counteract the surface tension caused by this curved surface 
the small curvature and keep a 
steady surface, the secondary high pressure must be required. 
The reverse flow from the separation point moves down along the jump
surface until it meets another high pressure (the primary high
pressure) on the upstream side of the jump.
The fluid motion is decelerated when it approaches the high pressure
on the inner side of the jump.
The direction of motion is then changed, and joins the main stream
again around the confluent point $R_{in}$.

\begin{figure}[hbtp]
  \begin{center}
  \psbox[height=1.5in]{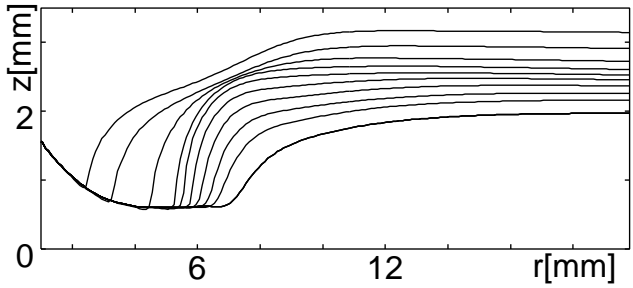}
  \end{center}
  \caption{Time evolution of the surface profile from the type II
           jump to the type I jump at $0.294$ s intervals.
           The topmost profile is the initial state.}  
  \label{fig:from2to1} 
\end{figure}

\begin{figure}[hbtp]
  \begin{center}
  \psbox[height=4.5in]{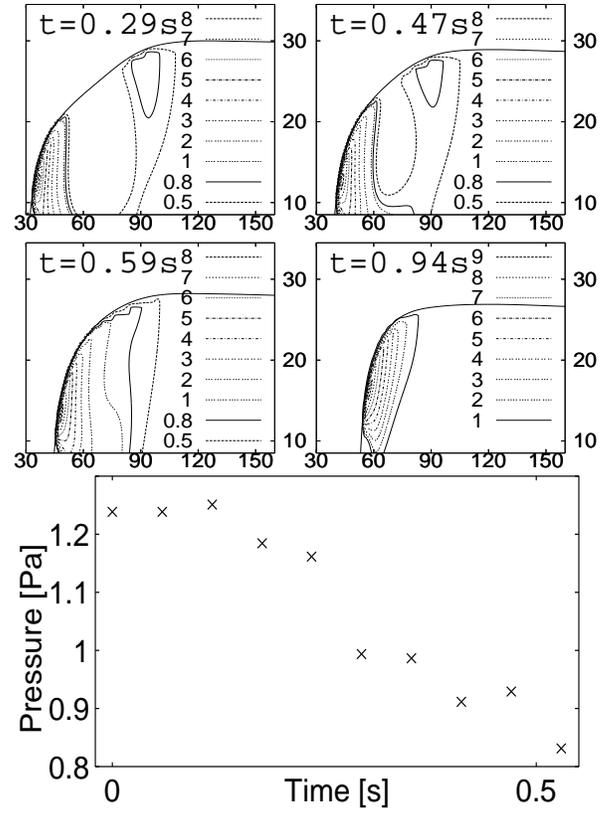}
  \end{center}
  \caption{Time evolution of the dynamic pressure distribution
           and the maximum value of the secondary high pressure.
           Fig.\ref{fig:pp} (c) and (a) correspond to $t=0$ and the
           final steady state.}  
  \label{fig:from2to1t2} 
\end{figure}

\begin{figure}[hbtp]
  \begin{center}
  \psbox[height=1.8in]{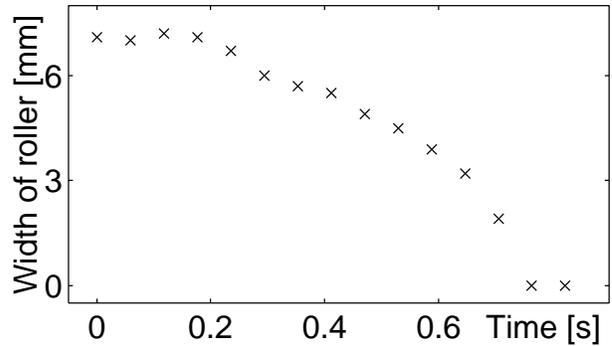}
  \end{center}
  \caption{Time evolution of the width of the roller.} 
  \label{fig:from2to1t1} 
\end{figure}

We further simulated the disappearance process of a roller (the
transition  
process from a type II jump to a type I jump) to study the details of 
the relation between the pressure field and the roller.  
We started from the steady state of a type II jump (the fourth
profile from the lowest in Fig.\ \ref{fig:result17}).
Its surface profile is shown as the topmost one in
Fig.\ \ref{fig:from2to1}.    
The time of this initial state was set $t=0$.
We simulated until the steady state of a type I jump (the second 
profile in Fig.\ \ref{fig:result17}) was reached by lowing the wall
height at $t=0$.   
Fig.\ \ref{fig:from2to1} displays the surface profiles at different
instants.
The flow experienced a transition from a type II jump to a type I
jump. 
The evolution of the dynamic pressure field and the maximum value of
the secondary high pressure are shown in Fig.\ \ref{fig:from2to1t2}.  
The initial pressure distribution is characterized by two
high pressure regions and a roller as discussed above.
As time increases, the secondary high pressure becomes weaker, and
finally vanishes around 0.55 s.
It appears that the reduction of the secondary high pressure is
associated with the decline in the curvature of the surface around the
secondary high pressure. 
Meanwhile, the primary high pressure does not experience any
significant change and finally becomes to the primary high pressure in
the type I jump.
In order to give a quantitative measure for the roller, we calculated
the horizontal width of the roller as $R_{out}^r-R_{in}^r$.
Fig.\ \ref{fig:from2to1t1} shows the time evolution of the roller
width $(R_{out}^r-R_{in}^r)$. 
With the secondary high pressure abating, the roller width decreased.
This process was significantly enhanced after the secondary high
pressure disappeared completely (from 0.55 s) because the pressure
gradient becomes perfectly opposite to the reverse flow of the roller.
Around 0.75 s, the roller disappeared.
With the secondary high pressure and the roller having abated, the
fluid eventually approached the steady state of a type I jump.

From this study, we have made clear that the existence of the high 
dynamic pressure regions and a secondary high pressure region around
the outer edge of a roller are essential to the maintenance of a
roller.
The establishment of the high pressure field is a result of the
balance among various fluid stresses, and the surface tension appears
to play an important role.
The secondary high pressure provides a driving force to generate a
reverse current beneath the jump surface in a type II jump.

We would like to thank S. Watanabe and K. Hansen for many discussions.
We also acknowledge the support of C. W. Stern.  
Numerical computations for this work were partially carried out
at the Computer Information Center, RIKEN
and the Yukawa Institute for Theoretical Physics, Kyoto University.

\end{document}